# Simplification of QPLDA:
# A practical method to the correction for the LDA band gap problem

Akihito Kikuchi

*Canon Inc., 30-2, Shimomaruko 3-chome, Ohta-ku, Tokyo 146-8501, Japan*

As an expedient way to correct the band-gap underestimation by LDA, Quasi-Particle-LDA (QPLDA) is proposed by several authors. Those preceding formalisms are not always efficient in practical calculations due to the lack of accuracy or the need of empirical parameters. The present work proposes a qualitatively accurate way of QPLDA which does not need any experimental parameters.

## I. INTRODUCTION

It is necessary to employ quasi-particle calculations to correct "band gap underestimation problems" in LDA. Since such methods are reliable but massive, several authors proposed Quasi-Particle-LDA (QPLDA) as an expedient way, where the total computational cost scales with the system size. Historically, Sham and Kohn introduced the idea of the "local mass operator" based on "local wavenumber" similar to WKB, but did not execute actual numerical calculations [1]. They took into account exchange interaction only, from which we could not expect precise treatments anyway. Later, Pickett and Wang had proposed a more qualitative method and had shown its potentiality in examples of semiconductors, such as silicon, diamond and GaP [2]. They used a model analytic formula for the dialectic function and adopted a model energy dispersion, which is free-electron like one, except that, being accompanied with an artificial band-gap discontinuity. The latter method has two shortcomings. First, to execute the calculation, several parameters are needed, such as the macroscopic dielectric constant and the band gap, which should be evaluated by the calculation itself, rather than being prepared as parameters. Second, in their computation of the mass operator, two or three dimensional numerical integrations are needed on each point of the direct space mesh, which demand large computational costs in iterative procedures in determining local wavenumbers and quasi-particle energies. This may be the reason why QPLDA has not been widely used until now. The present work points out that the slight modification to the original formalism of QPLDA by Sham and Kohn, without any ad-hoc parameters, can produce the better LDA band gap estimation.

## II. COMPUTATIONAL METHOD

At first, the basic QPLDA methodology is briefly reviewed. In QPLDA, the self energy operator is approximated as a local operator in the following expression, where $\varphi(r)$ is the classical electrostatic potential from electrons and nuclei.

$$\Sigma(r, r'; E) = \varphi(r)\delta(r - r') + \mathcal{M}(r, r', E - \varphi(r_0)); \quad r_0 = \frac{r + r'}{2}. \tag{1}$$

Thomas-Fermi relation in eq. (2) is applied to eliminate $\varphi(r)$.

$$\mu = \varphi(r) + \mu_h(n(r)) \tag{2}$$

This gives the relation between $\varphi(r)$, the crystal chemical potential $\mu$, and the "local" chemical potential $\mu_h$ (the chemical potential of a uniform gas of density $n(r)$ with $\varphi(r) = 0$). The mass operator becomes

$$\mathcal{M}_h(r, r', E - \varphi(r_0)) \approx \mathcal{M}_h\left(r, r', E - \mu + \mu_h(n(r_0))\right). \tag{3}$$

The non-local mass operator is approximated as a "local operator" operating on a single coordinate r (= r').

$$u(r; E) = \mathcal{M}_h[k_{LD}(r), E - \mu + \mu_h(n(r)); n(r)]. \tag{4}$$

To evaluate the mass operator, QPLDA uses plane-wave-like wavefunctions with local wavenumber $k_{LD}$, depending on the energy E and r.

$$\chi(r; E) \cong e^{ik_{LD}(r;E) \cdot r}. \tag{5}$$

The local wavenumber $k_{LD}$ is given by (in the wave space representation)

$$\frac{1}{2}k_{LD}^2 + \mathcal{M}_h(k_{LD}, E - \mu + \mu_h(n); n) = E - \mu + \mu_h(n). \tag{6}$$

If $k_{LD} = k_F$ ($k_F = (3\pi^2 n)^{1/3}$, the Fermi wavenumber) and $E = \mu$, $k_{LD}$ is given by

$$\frac{1}{2}k_F^2 + \mathcal{M}_h(k_F, \mu_h(n); n) = \mu_h(n). \tag{7}$$

The relationship between LDA exchange correlation potential $\mu_{xc}(n)$, the local mass operator

$u(r; \mu)$ and $\mu_h(n)$ is as follows.

$$u(r; \mu) = \left(\mu_h(n) - \frac{1}{2}k_F^2(n)\right) \equiv \mu_{xc}(n). \tag{8}$$

In the present work, in an analogy with the LDA formulation $\mu_{xc}^{LDA} = \mu_{exc}^{LDA} + \mu_{cor}^{LDA}$, the mass operator is separated into the exchange part and the correlation part.

$$\mathcal{M}_h(r; E; n) = \mathcal{M}_h^{exc}(r; E; n) + \mathcal{M}_h^{cor}(r; E; n). \tag{9}$$

As the exchange part of the mass operator, the expression by Hartree-Fock model is adopted, as used in ref.1,

$$u_{xk}(r) = -\frac{1}{\psi_k^*(r)\psi_k(r)} \sum_{k'=1}^{N} \int \frac{\psi_k^*(r)\psi_{k'}^*(r')\psi_{k'}(r)\psi_k(r')}{|r-r'|} dr'. \tag{10}$$

Assuming the plane-wave-like wavefunction, after the integration in the occupied states, we can obtain the analytic expression of the exchange part expressed by the local wavenumber and the Fermi wavenumber $k_F$.

$$u^{exc} = \mathcal{M}_h^{exc}(k_{LD}, E - \mu + \mu_h(n); n) = \mathcal{M}_h^{HF}(k_{LD}; n)$$
$$= -\frac{k_F}{\pi}\left[1 + \frac{k_F^2 - k_{LD}^2}{2k_{LD}k_F}\ln\left|\frac{k_F + k_{LD}}{k_F - k_{LD}}\right|\right]. \tag{11}$$

On the other hand, there is no such a simple analytic expression for the correlation part, whose contributions was neglected in ref.1. For this part, we approximate it by the correlation potential of LDA $\mu_{cor}(n)$, expressed by the local density.

$$\mathcal{M}_h^{cor}(r; E; n) \approx \mu_{cor}(n). \tag{12}$$

Consequently, $\mu_h(n)$ is given as

$$\mu_h(n) = \frac{1}{2} k_F^2 - \frac{k_F}{\pi} + \mu_{cor}(n). \tag{13}$$

The local wavenumber $k_{LD}$ is given by

$$\frac{1}{2} k_{LD}^2 + \mathcal{M}_h^{HF} + \mu_{cor}(n) = E - \mu + \mu_h(n). \tag{14}$$

After the substitution of $\mu_h(n)$, it becomes a non-linear equation,

$$\frac{1}{2} k_{LD}^2 + \mathcal{M}_h^{HF} = E - \mu + \frac{1}{2} k_F^2 - \frac{k_F}{\pi}. \tag{15}$$

Once E (the guess to the quasi-particle spectrum), $\mu$ and $k_F$ are given, $k_{LD}$ is determined as the solution in each point in the direct space. The value of the chemical potential $\mu$ is chosen to be the energy level of the LDA valence bands top. The Fermi wavenumber $k_F$ is also prepared by LDA calculations. We cannot always expect real-valued solutions for $k_{LD}$. If the real-valued solutions are not obtained, the logarithmic part of the mass operator should be replaced as

$$\ln \left| \frac{k_F + k_{LD}}{k_F - k_{LD}} \right| \rightarrow \ln \left( \frac{k_F + k_{LD}}{k_F - k_{LD}} \right). \tag{16}$$

It enables us to obtain the real valued mass operator even when $k_{LD}$ is a pure imaginary number.

The LDA energy spectrum $\varepsilon_j = \langle \psi_j | H_{LDA} | \psi_j \rangle$ is corrected by the next equation, whose solution of E is the quasi particle spectrum $E_{QP}$.

$$\begin{aligned} E &= \langle \psi_j | H_{LDA} - \mu_{xc}(n(r)) + \mathcal{M}_h(k_{LD}(r, E); n(r)) | \psi_j \rangle \\ &= \langle \psi_j | H_{LDA} - \mu_x(n(r)) + \mathcal{M}_h^{HF}(k_{LD}(r, E); n(r)) | \psi_j \rangle. \end{aligned} \tag{17}$$

## III. RESULT AND DISCUSSION

Three calculated examples are shown in Table 1, for diamond and BaTiO$_3$, in which the LDA band gaps at the Γ point are corrected by the present method. LDA calculations are executed by pseudopotential PW method, with the LDA parametrization of Perdew and Zunger [3]. In the preparing LDA calculation for BaTiO3, the partial core correction for the exchange-correlation is applied to Ti and Ba. In example 2, the QPLDA correction by the present work is executed by using full charge (valence + core), while, in example 3, the correction is executed by using valence charge only. In those cases, the LDA band gaps show the underestimation of about 60%, in comparison with the experimental value, while the present method improves the LDA band gaps to more reliable values, in which the deviations from the experiments is about 3%.

In fig.1, the QPLDA correction in BaTiO$_3$ is illustrated for the valence band maximum and the conduction band minimum at the Γ point. The vertical dotted lines are the positions of the LDA eigenvalues, and the broken lines are the expectation value of the Hamiltonian including QPLDA correction, i.e. $\langle i|H_{LDA} + \Delta_{QPLDA}(E)|i\rangle$ calculated at the energy E. The oblique real line illustrates that of y=x. Therefore the junctions of the real and the broken lines indicate quasi-particle energies. As the expectation value of the Hamiltonian shows an almost linear dependence on E, the secant method is an effective way in the determination of the spectrum.

In the comparison with former QPLDA formulations, the merit of the present method is summarized as follows. The mass operator in the present work is constructed from the exchange part and the correlation part. The exchange part is expressed by a Hartree-Fock type formula, which is the same as in ref. 1, and picks up many-body contributions. On the other hand, the correlation part, being neglected in ref. 1, is included in the present work, which is a qualitative improvement. Although the correlation part is merely approximated by the LDA correlation potential and the many-body correction is not applied [6], this treatment affords us a sufficient good result, because the necessity of the many-body correction in the correlation part is comparatively smaller than that in the dominating exchange part, as is suggested by the calculated examples. Furthermore, in our method, no ad-hoc parameters are needed, and the mass operator evaluation does not include time-consuming multi-dimensional numerical integrations as was employed in ref. 2, which are also advantages of the present work.

[6] If the many body correction should be necessary, the following trick may work: we can evaluate the "renormalized density" n*, thorough $\mu_{exc}^{LDA}(n^*) \equiv M_h^{HF}(k_{LD}; n)$ and evaluate the correlation term by $\mu_{cor}(n^*)$.

Table 1    The improvement of the direct band gap estimation by the present work.

|   |   | K point | LDA result | The present work | The experimental value |
|---|---|---|---|---|---|
| 1 | Diamond | Γ | 5.4eV | 7.6eV | 7.3eV[4] |
| 2 | BaTiO$_3$(Paraelectric) | Γ | 1.7eV | 3.1eV | 3.2eV[5] |
| 3 | BaTiO$_3$(Paraelectric) | Γ | 1.7eV | 3.4eV | 3.2eV[5] |

Figure Captions.

Fig. 1 QPLDA correction in BaTiO3.

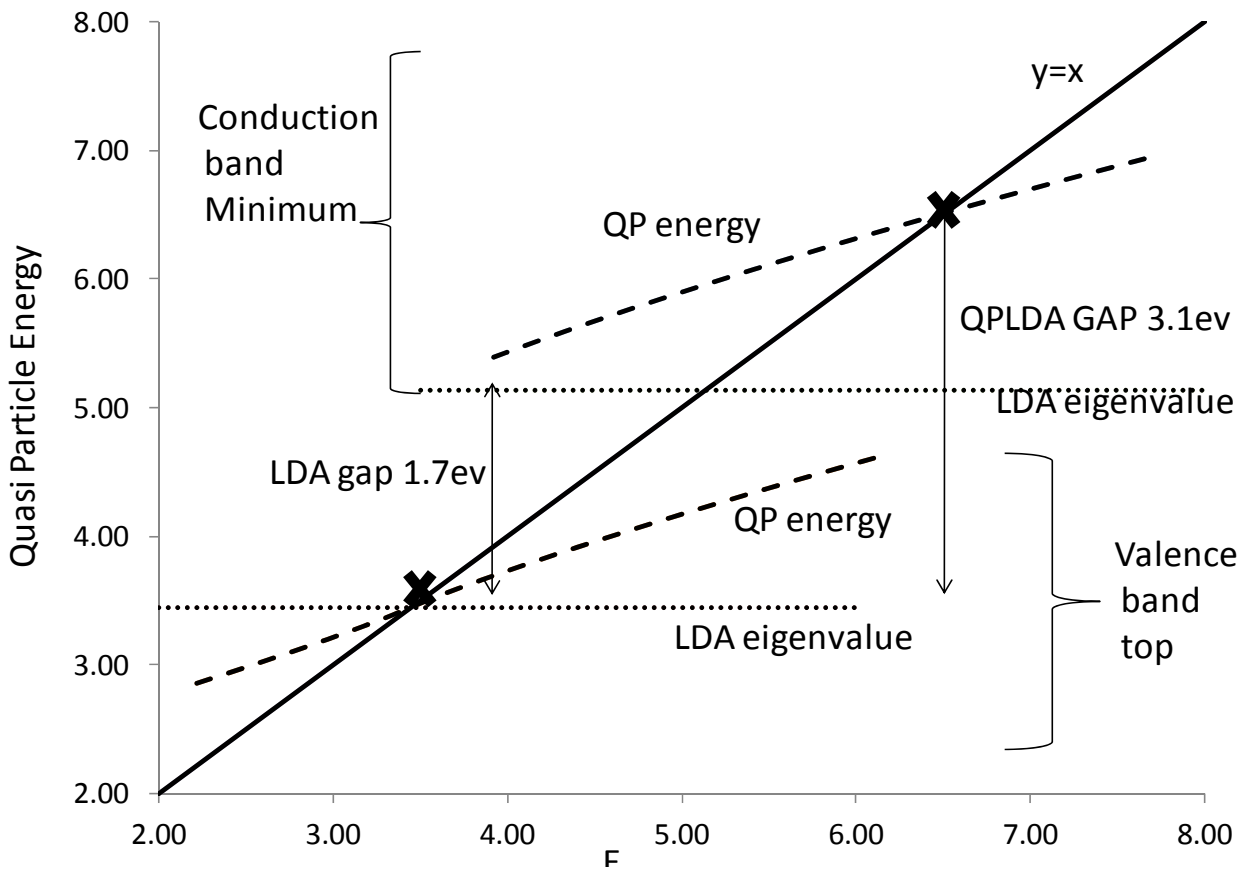

Fig.1